\documentstyle[12pt]{article}
\begin{document}

\begin{flushright}
SFU-HEP-105-93 \\
\end{flushright}
\vskip.3in

\begin{center}
 {\Large\bf Extrinsic Curvature Induced 2-d Gravity}\\
 \vspace{0.6in}

{\bf K.S.Viswanathan and R.Parthasarathy\footnote{Permanent
 address:The Institute of Mathematical Sciences, Madras 600 113, India.}}\\
\vspace{0.2in}
Department of Physics, Simon Fraser University\\
 Burnaby, B.C., Canada V5A 1S6\\
 \vspace{0.15in}
 Revised Version on \today
\end{center}
\vspace{0.3in}
 \begin{abstract}
 2-dimensional fermions are coupled to extrinsic geometry of a
 conformally immersed surface in ${\bf R}^3$ through gauge coupling. By
 integrating out the fermions, we obtain a WZNW action involving
 extrinsic curvature of the surface. Restricting the resulting effective
 action to surfaces of $h\sqrt g=1$, an explicit form of the action
 invariant under Virasaro symmetry is obtained. This action is a sum of
 the geometric action for the Virasaro group and the light-cone  action
 of 2-d gravity plus an interaction term. The central charges of the
 theory in both the left and right sectors are calculated.
 \end{abstract}
 \vspace{0.2in}
 \LaTeX Typesetting
 \newpage
\section{Introduction}
2-d gravity induced by a conformal field theory was first examined by
Polyakov \cite{1}. He found that there is an $Sl(2,R)$ current algebra.
This opened the way to solve off-critical string theory.
Knizhnik, Polyakov, and Zamolodchikov \cite{2} exploited this hidden
$Sl(2,R)$ symmetry to evaluate the scaling dimensions of planar random
surfaces. Their results showed complete agreement with the numerical
simulations  \cite{3}.

The geometrical origin of $Sl(2,R)$ is intriguing. Polyakov \cite{4}
demonstrated how one might get diffeomorphisms out of restricted
$Sl(2,R)$ gauge transformations. The mechanism resembles the phenomenon
of transformation of iso-spin into ordinary spin in the presence of
magnetic monopoles. It relies on background $Sl(2,R)$ gauge fields being
partially gauge fixed. This leaves a residual gauge symmetry which
solders iso-space and ordinary space yielding diffeomorphism. Extensions
to other groups and gauge restrictions have been shown to yield W-algebras.

The present authors have explored \cite{5} the possibility of realizing
gauge fields in an intrinsic manner which admits partial gauge fixing in
a geometrical context that leads to a theory of induced 2-d gravity. It
was demonstrated that if one takes into account the extrinsic geometry of
2-d surfaces in d- dimensions then one can construct $SO(d)$ gauge
fields from the components of the second fundamental form of the surface.
This observation is of course well known in the form of Gauss-Codazzi
equations of the immersed surface. What is interesting is that when we
consider surfaces with the property $h\sqrt{g}=1$ in the target space
${\bf R}^{3}$,where $h$ is the mean scalar curvature and $g$ is the
determinant of the induced metric on the surface,there arises a hidden
symmetry in the geometrical theory of such surfaces. This symmetry is
precisely the Virasaro symmetry. A similar property exists in ${\bf
R}^{4}$ also. Here the hidden symmetry is Virasaro$\oplus$Virasoro, when one
restricts to $h^{i}\sqrt{g}=1/\sqrt{2}$ (i=1,2), where $ h^{i}$ are the
two principal mean scalar curvatures. Similar conclusions were arrived at
in \cite{6} in the light cone gauge, whereas our results in \cite{5} are
in the conformal gauge. Beyond ${\bf R}^{4}$ it is not clear if there
exists a hidden symmetry for any special choice of surfaces, but
W-algebras have been obtained for surfaces immersed in affine spaces
\cite{7}.

It turns out to be extremely interesting to pursue further our
observations in \cite{5}. In this article we first obtain an action which
is explicitly invariant under Virasaro symmetry. Our considerations are
for surfaces in ${\bf R}^{3}$ but we see no difficulty in principle to
extending them to ${\bf R}^{4}$. This action depends on the components
$H_{\bar z\bar z}$ and $\frac{H_{zz}}{\sqrt{g}}$ of the second fundamental form
as the basic fields, the first one playing the role of an induced metric
while the second one that of the energy-momentum tensor. Actually, the
procedure for obtaining this action was worked out by Polyakov in
\cite{4}. The resulting action is a WZNW action. One finds that the
effective theory is described by a sum of the geometric action for the
Virasaro group \cite{8}, the action for 2-d gravity in the light cone
gauge \cite{2}, and an interaction term which is needed to make the
action invariant under the Virasaro transformation. We then analyze the
properties of the effective action under conformal transformations in the
left and the right sectors and compute their central charges for the
quantum theory of the extrinsic geometry. We find that the left sector is
described by an energy momentum tensor $T_{\bar z\bar z}$ which is a sum of
$T_{\bar z\bar z}$ of the light cone action and $T_{\bar z\bar z}$ of the
geometric
action. They are both of the modified Sugawara form. The right
sector, however, has only Virasaro symmetry and the energy-momentum tensor
$T_{\bar z\bar z}$ is that of the geometric action only. Luckily, the central
charges for both these cases are known. Recently, Aoyoma \cite{9} has
calculated the central charges $c_{L,R}$ of the geometric action, while
$c_{L}$ of the light cone action is known from \cite{1}, and \cite{2}. After a
careful consideration of the contribution of the ghosts in both
sectors, we find that the theory of induced 2-d gravity induced from the
extrinsic geometry in ${\bf R}^{3}$ works for conformal matter fields
with $d>1$ in both sectors. It has been remarked in \cite{4} that the
main stimulus for finding the gauge representation arises from the hope
of describing $d>1$ conformal fields in 2-d quantum gravity. Our
results suggest that extrinsic curvature does indeed become fundamental
variables in the case $d>1$, rather than the intrinsic metric of the
surface.

\section{Basic properties of $h g^{1/2}=1$ surfaces}
Although this section contains no new results, we feel that it is useful
to recall the basic properties of $h\sqrt{g}=1$ surfaces conformally
immersed in ${\bf R}^{3}$ so that this article is self-contained. For
details we refer the reader to \cite{5}.

The second fundamental form of a surface in ${\bf R}^{d}$ is defined by
\begin{equation}
\partial_{\alpha}\partial_{\beta}X^{\mu}=
\Gamma_{\alpha \beta}^{\gamma}\partial_{\gamma}X^{\mu} +
H^{i}_{\alpha \beta}N_{i}^{\mu},
\label{2.1}
\end{equation}
where $X^{\mu}(\xi^{\alpha})$ are the string
coordinates, $\Gamma_{\alpha\beta}^{\gamma}$ the affine connection defined
by the induced metric
$g_{\alpha\beta}=\partial_{\alpha}X^{\mu}\partial_{\beta}X^{\mu}$ and
$N_{i}^{\mu} (i=1...d-2)$ are the normals to the string world sheet. We
treat both the world sheet and the target space as Euclidean. The normals
$N_{i}^{\mu}$ satisfy the following equations,
\begin{equation}
\partial_{\alpha}N_{i}^{\mu}=\tau^{ij}_{\alpha}N^{\mu}_{j}
-H^{i\gamma}_{\alpha}\partial_{\gamma}X^{\mu},
\label{2.2}
 \end{equation}
 where $\tau^{ij}_{\alpha}(\alpha=1,2)$ are the normal connections of the
 surface. (1) and (2) are also referred to as the Gauss-Codazzi
 equations. The scalar curvature R of the world sheet is given as
 \begin{equation}
 	R=(H^{i\alpha}_{\alpha})^{2}
 	-H^{i\beta}_{\alpha}H^{i\alpha}_{\beta}.
 	\label{2.3}
 	\end{equation}
 	It is convenient to rewrite the structure equations (\ref{2.1})
 	and (\ref{2.2}) as follows;
 	\begin{equation}
 		\partial_{z}{\bf\hat e{_{i}}}=(A_{z})_{ij}{\bf \hat e{_{j}}}
 		\ \ \ \ (i,j=1,2,3.).
 		\label{2.4}
 		\end{equation}
 A similar equation holds for the $\bar z$ derivative. Here
 $z=\xi_{1}+i\xi_{2}$ and $\bar z=\xi_{1}-i\xi_{2}$ are the local
 isothermal coordinates on the surface. $\bf\hat e_{i}$ is a local
 orthonormal moving frame in which $\bf\hat e_{1,2}$ are tangential and
 $\bf\hat e_{3}$ is normal to the surface. $(A_{z})_{ij}$ is antisymmetric in
 (ij) whose elements are given in terms of $H^{i}_{\alpha\beta}$. It
 transforms under local $SO(3)$ gauge transformations
 		\begin{equation}
 			\hat{\bf e}'_{i}=g_{ij}\hat{\bf e}_{j}
   \label{2.5}
          \end{equation}
          as a gauge field. The gauge transformations generate
          $\bf\hat{e}'_{i}$ which form a local moving frame to a different
          surface on which the induced metric is again in the conformal
          gauge. The two tangent vectors $\hat{\bf e}_{1}$ and $\hat{\bf
          e}_{2}$ to the surface are determined only upto an $SO(2)$
          rotation. It is possible to choose these tangent vectors in
          such a way that $(A_{z})_{12}=0$. The explicit forms of $A_{z}$
          and $A_{\bar z}$ after this rotation are given in [5]. If we
          now restrict to surfaces of $h\sqrt{g}=1$ then the residual
          gauge transformation is a diffeomorphism. Of the three complex
          parameters determining $SO(3,C)$, two are determined in terms of
          the third. Then we find that $A^{+}_{z}\equiv
          T_{zz}=\frac{H_{zz}}{\sqrt{g}}$ transforms as the energy-
          momentum tensor
          \begin{equation}
          	\delta_{\epsilon^{-}}T_{zz}=-\partial_{z}^{3}\epsilon^{-}
          	-2\partial_{z}\epsilon^{-}T_{zz}-\epsilon^{-}\partial_{z}T_{zz}
          	\label{2.6}
          	\end{equation}
          	$\frac{H_{zz}}{\sqrt{g}}$ thus plays the role of energy
          	momentum tensor for the extrinsic geometry induced 2-d
          	gravity. An analysis of the transformation of $A_{\bar z}$
          	shows that $A_{\bar z}^{-}\equiv H_{\bar z\bar z}$ transforms
          	as the metric tensor
          	\begin{equation}
          		\delta_{\epsilon^{-}}H_{\bar z\bar z}=\partial_{\bar
          		z}\epsilon^{-}+\epsilon^{-}\partial_{z}H_{\bar z\bar
          		z}-(\partial_{z}\epsilon^{-})H_{\bar z\bar z}
          		\label{2.7}
          		\end{equation}

          		Finally the integrability condition which amounts to saying
          		that $F_{z\bar z}=0$ ( $(A_{z},A_{\bar z})$ is a pure gauge)
          		leads to the anomaly equation of 2-d gravity:
          		\begin{equation}
          			\partial_{z}^{3}H_{\bar z\bar z}=\partial_{\bar z}
          			T_{zz}-2(\partial_{z}H_{\bar z\bar z})T_{zz}
          			-H_{\bar z\bar z}\partial_{z}T_{\bar z\bar z}
          			\label{2.8}
          			\end{equation}
          			Similar results have been derived for surfaces in ${\bf
          			R}^{4}$
          			[5] but we shall not need them here as we restrict to
          			${\bf R}^{3}$ in the sections below.

\section{WZNW action for extrinsic geometry}
This section contains results derived in \cite{4} in the context
of diffeomorphisms from gauge transformations. We need only to
identify the gauge field with appropriate components of the
extrinsic curvature tensor. For completeness and continuity we
sketch this procedure.
   We wish to derive the WZNW action that exhibits explicitly
the Virasaro symmetry in $h\sqrt g = 1$ surfaces. One couples
the gauge fields discussed in the previous section to 2-d
fermions in a gauge invariant manner and then integrate out the
fermions. The resulting action is known to be a WZNW action and
can be written as,
\begin{eqnarray}\label{3-1}
{\Gamma}_{eff}={\Gamma}_{-}(A_z) + {\Gamma}_{+}(A_{\bar{z}})
- Tr \int A_z A_{\bar{z}} d^2\xi,
\end{eqnarray}
where $A_z=h^{-1}{\partial}_z h$, and
$A_{\bar{z}}=g^{-1}{\partial}_{\bar{z}} g$, $h,g \in SO(3,C)$.
We next derive the form of (\ref{3-1}) under the gauge
restriction $A^-_z=1(h\sqrt g=1)$ and $A^0_z=0$. Then,
\begin{eqnarray}
{\Gamma}_-(A^-_z=1,A^0_z=0,A^+_z\equiv \frac{H_{zz}}{\sqrt
g})&\equiv & S_-(F_1), \nonumber
\end{eqnarray}
can be  explicitly constructed by parameterizing $A^+_z$ as,
\begin{eqnarray}\label{3-2}
A^+_z\equiv \frac{H_{zz}}{\sqrt g}&=&D_zF_1 \nonumber \\
&\equiv & \frac{{\partial}^3_zF_1}{{\partial}_zF_1} -
\frac{3}{2}
{\left(\frac{{\partial}^2_zF_1}{{\partial}_zF_1}\right)}^2,
\end{eqnarray}
to be,
\begin{eqnarray}\label{3-3}
S_-(F_1)=\frac{1}{2}\int
\left(\frac{{\partial}_{\bar{z}}F_1}{{\partial}_zF_1}\right)
\left[\frac{{\partial}^3_zF_1}{{\partial}_zF_1}-2{\left(
\frac{{\partial}^2_zF_1}{{\partial}_zF_1}\right)}^2\right]
dz\wedge d\bar{z}.
\end{eqnarray}
What is appealing about this action is that the field $F_1$ has
a geometrical interpretation. It is easily derived from our
results \cite{5} on Gauss map of 2-d surfaces into the Grassmannian
manifold that $F_1$  in (\ref{3-3}) is the $CP^1(\simeq G_{2,3})$
field that arises in the Gauss map. (\ref{3-3}) is the geometric
action for the Virasaro group studied by Alekseev and
Shatashvili \cite{8}.
   The quantum action $S_{+}(H_{\bar{z}\bar{z}})$ is defined by,
\begin{eqnarray}\label{3-4}
exp(-S_{+}(H_{\bar{z}\bar{z}})) = \int [dA_{\bar{z}}]\delta
(A^-_{\bar{z}}-H_{\bar{z}\bar{z}})exp(-({\Gamma}_{+}(A_{\bar{z}})-\int
A^{+}_{\bar{z}})).
\end{eqnarray}
In the classical limit, evaluating the path integral at the
stationary point,
\begin{eqnarray}\label{3-5}
\frac{\delta {\Gamma}_{+}}{\delta A^0_{\bar{z}}}\equiv J^0_z=0
&;&\frac{\delta {\Gamma}_{+}}{\delta A^{+}_{\bar{z}}}\equiv
J^-_z=1,
\end{eqnarray}
we have,
\begin{eqnarray}\label{3-6}
S^{cl}_{+}={\min_{A^0_{\bar{z}},A^{+}_{\bar{z}}}} \left[ {\Gamma}_{+}
(A_{\bar{z}})-\int A^{+}_{\bar{z}}\right].
\end{eqnarray}
It is interesting to note that from our discussion in section 2,
$A^-_{\bar{z}}$ after gauge rotation is indeed given by
$H_{\bar{z}\bar{z}}$, where $H_{\bar{z}\bar{z}}$ is the component of the
second fundamental form. It is shown in \cite{4} that (\ref{3-6}) may be given
an explicit form by,
\begin{eqnarray}\label{3-7}
S^{cl}_{+}(F_2)=-\frac{1}{2}\int \left[
({\partial}^2_zF_2)\frac{{\partial}_z{\partial}_{\bar{z}}F_2}
{({\partial}_zF_2)^2} - ({\partial}^2_zF_2)^2\frac{{\partial}_{\bar{z}}F_2}
{({\partial}_zF_2)^3}\right] dz\wedge d\bar{z}.
\end{eqnarray}
This is precisely the light-cone gauge action of 2-d gravity \cite{1,2}. Indeed
$J^-_z=1$ and $J^0_z=0$ play the role of light cone gauge fixing
conditions on the metric, namely, $h_{--}=0$, $h_{-+}=1$. In (\ref{3-7}),
$F_2$ is related to $H_{\bar{z}\bar{z}}$ by,
\begin{eqnarray}\label{3-8}
H_{\bar{z}\bar{z}}=\frac{{\partial}_{\bar{z}}F_2}{{\partial}_zF_2}.
\end{eqnarray}
The complete action (\ref{3-1}) is thus given by,
\begin{eqnarray}\label{3-9}
{\Gamma}^{eff}(F_1,F_2)&=&\frac{1}{2}\int
\frac{{\partial}_{\bar{z}}F_1}{{\partial}_zF_1}\left[\frac{{\partial}^3_zF_1}
{{\partial}_zF_1}-2\frac{({\partial}^2_zF_1)^2}{({\partial}_zF_1)^2}
\right] \nonumber \\
&-&\frac{1}{2}\int
\left[{\partial}^2_zF_2\frac{{\partial}_z{\partial}_{\bar{z}}F_2}
{({\partial}_zF_2)^2}-({\partial}^2_zF_2)^2\frac{{\partial}_{\bar{z}}F_2}
{({\partial}_zF_2)^3} \right] \nonumber \\
&-&\int \frac{{\partial}_{\bar{z}}F_2}{{\partial}_zF_2}D_zF_1.
\end{eqnarray}
${\Gamma}^{eff}$ can be shown to be explicitly invariant under
transformations (\ref{2.6}) and (\ref{2.7}) of $\frac{H_{zz}}{\sqrt g}$ and
$H_{\bar{z}\bar{z}}$ respectively. Thus (\ref{3-9}) is the extrinsic
curvature induced 2-d gravity action that exhibits Virasaro symmetry
claimed in \cite{5}. It is worth remarking that this action contains both the
geometric and the light-cone actions with a gauge invariant coupling.
Incidentally, the last term in (\ref{3-9}) may be written as $\int {\mid
H\mid}^2\sqrt g d^2\xi $ which is precisely the extrinsic curvature
action. The equation of motion following from (\ref{3-9}) is,
\begin{eqnarray}\label{3-10}
{\bigtriangledown}_{\bar{z}}(\frac{H_{zz}}{\sqrt g})={\partial}^3_z
H_{\bar{z}\bar{z}},
\end{eqnarray}
where,
\begin{eqnarray}\label{3-11}
{\bigtriangledown}_{\bar{z}}={\partial}_{\bar{z}}-2({\partial}_zH_
{\bar{z}\bar{z}})-H_{\bar{z}\bar{z}}{\partial}_z.
\end{eqnarray}
The main result then is that we have an action described by the fields
$H_{\bar{z}\bar{z}}=\frac{{\partial}_{\bar{z}}F_2}{{\partial}_zF_2}$ and
$\frac{H_{zz}}{\sqrt g}=D_zF_1$, which has Virasaro symmetry. This action
is basically a superposition of the geometric and the light-cone 2-d
gravity actions and hence is the statement that extrinsic geometry
induces an effective 2-d gravity theory.

\section{Properties of ${\Gamma}_{eff}$ under conformal transformations}

   Following Aoyama [9] we assign conformal weights
$({\Delta}_{\bar{z}},{\Delta}_z) = (0,0)$ to the fields $F_1$ and
$F_2$. In the $\bar{z}$ sector, the conformal transformations are
given by,
\begin{eqnarray}\label{4-1}
\delta^{\bar{z}}_{conf}
F_i=\bar{\epsilon}(\bar{z}){\partial}_{\bar{z}}F_i,\ \ \ (i=1,2).
\end{eqnarray}
Computing the change in ${\Gamma}_{eff}$ under (\ref{4-1}), we find,
\begin{eqnarray}\label{4-2}
\delta{\Gamma}_{eff}&=&-\int \bar{\epsilon}(\bar{z})\left\{
\frac{{\partial}_{\bar{z}}F_1}{{\partial}_zF_1}
{\partial}_{\bar{z}}(D_zF_1) +
\frac{{\partial}_{\bar{z}}F_2}{{\partial}_zF_2}
{\partial}^3_z(\frac{{\partial}_{\bar{z}}F_2}{{\partial}_zF_2})\right\}
\nonumber \\
&\equiv & -\int \bar{\epsilon}(\bar{z}){\partial}_z T_{\bar{z}\bar{z}},
\end{eqnarray}
where,
\begin{eqnarray}\label{4-3}
T_{\bar{z}\bar{z}}&\equiv & T_{\bar{z}\bar{z}}(F_1) +
T_{\bar{z}\bar{z}}(F_2) \nonumber \\
&=&-\frac{1}{2}\left\{  {\left(
\frac{{\partial}_z{\partial}_{\bar{z}}F_1}{{\partial}_zF_1}\right)}^2 -
2({\partial}_{\bar{z}}F_1)\left(
\frac{{\partial}^2_z{\partial}_{\bar{z}}F_1}{({\partial}_zF_1)^2} -
\frac{({\partial}^2_zF_1){\partial}_z{\partial}_{\bar{z}}F_1}{({\partial}_z
F_1)^3}\right)\right\} \nonumber \\
&+&\left\{ \frac{{\partial}_{\bar{z}}F_2}{{\partial}_zF_2}
{\partial}^2_z\left(
\frac{{\partial}_{\bar{z}}F_2}{{\partial}_zF_2}\right) -
\frac{1}{2}{\left({\partial}_z\frac{{\partial}_{\bar{z}}F_2}{{\partial}_zF_2}
\right)}^2 \right\}.
\end{eqnarray}
It is interesting to note that the interaction term in (\ref{3-9}) is
invariant under conformal transformations in the $\bar{z}$
sector. $T_{\bar{z}\bar{z}}(F_1)$ agrees with the expression in
[9] where it is shown that it is of the Sugawara form for
$S\ell(2,R)$ currents. $T_{\bar{z}\bar{z}}(F_2)$ agrees with the
expression in KPZ[2]. Thus the total energy momentum tensor in
the $\bar{z}$ sector is a sum of the Sugawara forms of
$T_{\bar{z}\bar{z}}$ for the geometric $(F_1)$ and the light
cone $(F_2)$ actions.

    In the $z$-sector the conformal transformations are given by,
\begin{eqnarray}\label{4-4}
{\delta}^z_{conf}F_i= \epsilon(z){\partial}_zF_i,\ \ \ (i=1,2).
\end{eqnarray}
Computing the change in the action, we find,
\begin{eqnarray}\label{4-5}
\delta  {\Gamma}_{eff}=-\int \epsilon(z)
{\partial}_{\bar{z}}(D_zF_1).
\end{eqnarray}
It is important to note that the energy momentum tensor in the
$z$-sector is,
\begin{eqnarray}\label{4-6}
T_{zz}=D_zF_1.
\end{eqnarray}
The variation of $S(F_2)$ cancels that of the interaction term
in (\ref{3-9}). This result agrees with the interpretation earlier in
section 2, of $D_zF_1\ =\ \frac{H_{zz}}{\sqrt g}$ as the energy
momentum tensor for the Virasaro symmetry. Thus to summarize,
we have found that the effective 2-d gravity induced by the
extrinsic geometry has Virasaro symmetry in the $z$-sector and
Virasaro-Kac-Moody symmetry in the $\bar{z}$-sector.

\section{Quantum properties of ${\Gamma}_{eff}$}

   In the $\bar{z}$ sector, we found that the total energy
momentum tensor of our theory is a sum of that of the geometric
action and that of the light-cone action. The geometric action
in this sector as well as in the $z$ sector has been quantized
by Aoyoma \cite{9}, while the light-cone action has been quantized
by KPZ \cite{2}. We can therefore make use of their results to
describe the quantum theory of extrinsic curvature induced 2-d
gravity in the $\bar{z}$ sector.

   The quantization of the light-cone action in the $z$ sector
has not been done so far. Fortunately, as we found above, the
energy momentum tensor of our theory in the $z$ sector is
identical to that of the geometric action given by $D_zF_1$,
and as remarked above this sector of the geometric action has
been quantized in \cite{9}.

   Let us discuss the $\bar{z}$ sector first. In the quantum
theory of the geometric action the conformal weights of the
field $F_1$ is changed in accordance with the observation that
${\partial}_zF_1\ =\ exp(\sqrt{\frac{2}{k}}\phi)$, tells us
that it should be $(1,0)$, since in the classical Liouville
theory, $exp(\sqrt{\frac{2}{k}}\phi)$ has conformal weight
$(1,1)$. Accordingly, we take,
\begin{eqnarray}\label{5-1}
{\delta}^{\bar{z}}_{conf}F_1=\epsilon(\bar{z}){\partial}_{\bar{z}}F_1
+ ({\partial}_{\bar{z}}\bar{\epsilon}(\bar{z}))F_1.
\end{eqnarray}
It is easily shown \cite{9} that $T'_{\bar{z}\bar{z}}(F_1)$ is given
by,
\begin{eqnarray}\label{5-2}
T'_{\bar{z}\bar{z}}(F_1)=T_{\bar{z}\bar{z}}(F_1) -
{\partial}_{\bar{z}}J^0_{\bar{z}},
\end{eqnarray}
where $T_{\bar{z}\bar{z}}(F_1)$ is defined in (\ref{4-3}) and,
\begin{eqnarray}\label{5-3}
J^0_{\bar{z}}=-\frac{1}{2}\left\{F_1\frac{{\partial}^2_z{\partial}_{\bar{z}}F_1}
{({\partial}_zF_1)^2} -
({\partial}^2_zF_1)\frac{{\partial}_z{\partial}_{\bar{z}}F_1}{({\partial}_zF_1)
^3} -
\frac{{\partial}_z{\partial}_{\bar{z}}F_1}{{\partial}_zF_1}\right\}.
\end{eqnarray}
It is known from \cite{9} that the central charge of this energy
momentum tensor is,
\begin{eqnarray}\label{5-4}
c^{\bar{z}}(F_1)=15 - 6(k+2) - \frac{6}{k+2}.
\end{eqnarray}
The results of KPZ \cite{2} can be used to find the central charge of
$T_{\bar{z}\bar{z}}(F_2)$. Recall that $
S^{cl}_{+}(H_{\bar{z}\bar{z}})$ has been obtained by imposing
the conditions (13), which can be identified with the
light-cone gauge conditions $h_{--}\ =\ 0$ and $h_{+-}\ =\ 1$
in equation 6 of \cite{2}. Thus the improved energy momentum tensor
is as given by KPZ \cite{2}. The central charge of this theory is
found to be same as (\ref{5-4}),
\begin{eqnarray}
c^{\bar{z}}(F_2)=c^{\bar{z}}(F_1). \nonumber
\end{eqnarray}

   Let us examine the ghost sector of our theory. Considering
first the geometric action, we recall that it was obtained by
the choice $ A^-_{z} \equiv h\sqrt g = 1$, and $A^0_{z} = 0$.
In the quantum theory they manifest as gauge restrictions and
by identifying $A^-_{z}=1$ with $h_{+-}=1$ and $A^0_{z}=0$ with
$h_{--}=0$ gauge choice in 2-d gravity, we find that the ghosts
contribute $-28$ to the central charge. The argument above
suggest that the ghost contribution to the central charge for
the light-cone action is also $-28$. Denoting the central
charge of the matter by $d$, we thus find,
\begin{eqnarray}\label{5-5}
c^{\bar{z}}_{tot}= d - 56 + 2( 15 - 6(k+2) - \frac{6}{k+2}).
\end{eqnarray}
By equating this to zero, we find,
\begin{eqnarray}\label{5-6}
k + 2 =\frac{d - 26 \pm ((2-d)(50-d))^{\frac{1}{2}}}{24}.
\end{eqnarray}
Because of the doubling of the gravity and ghost contributions,
we find that the theory of induced  2-d gravity makes sense for
$d \leq 2 $.

   Turning now to the $z$-sector, our theory is equivalent to
the $z$-sector of the geometric action. From the result of
Aoyoma \cite{9} we have,
\begin{eqnarray}\label{5-7}
c_z = 13 - 6(k+2) - \frac{6}{k+2}.
\end{eqnarray}
The ghost contribution is once again -28. Thus, we have,
\begin{eqnarray}\label{5-8}
c^{z}_{tot}= d - 28 + 13 -6(k+2) - \frac{6}{k+2}.
\end{eqnarray}

{}From (\ref{5-8}) we find,
\begin{eqnarray}
k + 2 = \frac{d - 15 \pm ((3-d)(27-d))^{\frac{1}{2}}}{12}.
\end{eqnarray}
{}From this we note that the theory in $z$-sector works for $d
\leq 3$.

\section{Summary}

   We have shown that the extrinsic curvature of immersed
surfaces with the property $h\sqrt g = 1$ induce an effective
2-d gravity theory described by a sum of the geometric and
light cone actions in a gauge invariant way. The central
charges in both the left and right sectors are calculable and
the theory makes sense for $d \geq 1$ in both the sectors.

\begin{center}
{\large\bf Acknowledgements}
\end{center}

   This work has been supported in part by an operating grant from
Natural Sciences and Engineering Council of Canada. We thank Zheng Huang,
T.Jayaraman, N.D.Haridass and V.John for valuable comments. One of the authors
(R.P)
thanks Drs.S.Wadia,G.Date and S.Govindarajan for useful discussions.
K.S.V.\  wishes to thank Matscience for the kind hospitality during his stay
at that Institute. R.P wishes to thank the Physics Department, Simon Fraser
University for kind hospitality.

 \end{document}